# Comparative study of forward and backward test-kinetic simulation approaches

Gabriel Voitcu[a,b], Marius Echim[a,c] and Richard Marchand[d]

[a]*Institute of Space Science, Atomistilor 409, Magurele 077125, Romania*

[b]*University of Bucharest, Faculty of Physics, Atomistilor 405, Magurele 077125, Romania*

[c]*Belgian Institute for Space Aeronomy, Avenue Circulaire 3, Brussels B-1180, Belgium*

[d]*University of Alberta, Department of Physics, T6G 2G7 Edmonton, Alberta, Canada*

Corresponding author: Gabriel Voitcu (gabi@spacescience.ro)

## Abstract

In this paper we perform a comparative study of the forward and backward Liouville mapping applied to the modeling of ring-shaped and non-gyrotropic velocity distribution functions of particles injected in a sheared electromagnetic field. The test-kinetic method is used to compute the velocity distribution function in various areas of a proton cloud moving in the vicinity of a region with a sharp transition of the magnetic field and a non-uniform electric field. In the forward approach the velocity distribution function is computed for a two-dimensional spatial bin, while in the backward approach the distribution function is averaged over a spatial bin with the same size as for the forward method and using a two-dimensional trapezoidal integration scheme. It is shown that the two approaches lead to similar results for spatial bins where the velocity distribution function varies smoothly. On the other hand, with bins covering regions of configuration space characterized by sharp spatial gradients of the velocity distribution function, the forward and backward approaches will generally provide different results.

## 1. Introduction

Test-kinetic simulations provide a useful tool to investigate the dynamics of charged particles in systems in which a good approximation of the actual electromagnetic fields can be obtained [1]. This approach can provide useful information about the kinetic structure of the system. The electric and magnetic fields used in the test-kinetic approach are prescribed a priori. Thus, the test-kinetic method gives a first approximation of the kinetic structure of a plasma using electric and magnetic fields obtained from either theoretical models, MHD simulations or experimental data. Although the results obtained using this approach are not self-consistent, the test-kinetic method is an important simulation tool able to provide a useful description of complex situations where the use of self-consistent kinetic methods is not possible.

The test-kinetic method has been applied in various contexts of space plasma physics. For example, Wagner et al. [2] integrated test-particle orbits in an X-line distribution of the magnetic field illustrating the non-adiabatic character of orbits for sheets having the thickness comparable with the particle's Larmor radius. Speiser et al. [3] used the test-kinetic approach to map velocity distribution functions from the magnetosphere into the magnetosheath. Curran et al. [4] and Curran and Goertz [5] mapped velocity distribution functions along numerically integrated orbits in order to study the plasma dynamics in X-line magnetic topology. Ashour-Abdalla et al. [6] investigated the ion dynamics into the magnetotail using the test-kinetic approach. Richard et al. [7] studied the magnetospheric penetration mechanisms by solar wind ions using the same method, for electric and magnetic field profiles obtained from a global MHD simulation of the terrestrial magnetosphere. Rothwell et al. [8] developed test-particle simulations in order to investigate non-adiabatic effects introduced by sharp spatial variations of the electromagnetic field. Delcourt et al. [9,10,11] performed test-particle simulations to investigate impulsive changes of ion dynamics in the near-Earth plasma sheet. Mackay et al. [12] and Marchand et al. [13] applied the test-kinetic approach to obtain first order kinetic effects in collisionless perpendicular shocks in the vicinity of the Earth's bow shock and also to check consistency with a solution obtained in the MHD approximation.

The test-kinetic simulation method is based on numerical integration of test-particle orbits in prescribed electric and magnetic fields. Marchand [1] identified four different approaches of the test-kinetic method: (i) trajectory sampling, (ii) forward Monte Carlo, (iii) forward Liouville and (iv) backward Liouville. In this paper we perform a comparative study of the forward and backward Liouville approaches in the test-kinetic modelling of ring-shaped and non-gyrotropic velocity distribution functions for particles injected in a sheared electromagnetic field. The test-kinetic simulation method is used to compute the velocity distribution function in various regions of a proton cloud moving in the vicinity of a region with a sharp transition of the magnetic field. This type of configuration is of interest for studying the dynamics of the terrestrial magnetotail. The distribution functions obtained with both approaches are compared and the differences are analyzed.

The remainder of the paper is organized as follows. In the second section we outline the main features of both forward and backward test-kinetic approaches and describe how they are applied in our simulations. In the third section we illustrate the numerical solutions obtained for a cloud of protons injected in a non-uniform electromagnetic field and we analyze the differences between the velocity distribution functions given by both approaches. The last section includes our conclusions.

## 2. Test-kinetic modeling: forward and backward Liouville approaches

In a collisionless plasma the characteristics of the Vlasov equation can be obtained by solving the Newton-Lorentz equation of motion:

$$\frac{d^2\vec{r}}{dt^2} = \frac{q}{m}\left( \vec{E} + \frac{d\vec{r}}{dt} \times \vec{B} \right) \tag{1}$$

for an ensemble of charged particles injected into the electromagnetic field given by $\vec{E}$ and $\vec{B}$ [14]. This is equivalent to Liouville's theorem applied to a one-particle distribution function which states that:

$$\frac{df}{dt} = 0 \tag{2}$$

along a particle trajectory. Therefore, the velocity distribution function has the same numerical value at each point along a particle orbit. One can compute any number of Vlasov characteristics and then "propagate" $f$ along them by applying Liouville's theorem.

In the forward and backward test-kinetic approaches the magnetic and electric fields introduced in Eq. (1) are prescribed. The B-field used in our computations is stationary and it varies with the $x$-coordinate in a transition region centred at $x$=0. The profile of the magnetic field is anti-parallel, i.e. $\vec{B}$ is everywhere parallel to the $z$-axis and it changes orientation as it goes through $x$=0:

$$B_z(x) = -B_{1z} \, \text{erf}\left(\frac{x}{L}\right) \tag{3}$$

where $B_{1z}$ represents the asymptotic field in the left hand side of the transition region ($x \to -\infty$), $-B_{1z}$ is the asymptotic field in the right hand side ($x \to +\infty$) and $L$ represents the characteristic scale length of the transition region. This type of magnetic profile has been obtained self-consistently from kinetic models of one-dimensional tangential discontinuities [15,16,17].

The electric field is everywhere normal to the magnetic induction $\vec{B}$ and it is obtained by solving the two-dimensional Laplace equation:

$$\frac{\partial^2 \Phi}{\partial x^2} + \frac{\partial^2 \Phi}{\partial y^2} = 0 \tag{4}$$

in the $xOy$ plane. The integration domain considered is rectangular with $x_L \leq x \leq x_R$ and $y_B \leq y \leq y_T$. The boundary conditions are taken to be Neumann, with:

$$\left.\frac{\partial \Phi}{\partial x}\right|_{x=x_L} = \left.\frac{\partial \Phi}{\partial x}\right|_{x=x_R} = 0$$
$$\left.\frac{\partial \Phi}{\partial y}\right|_{y=y_B} = \left.\frac{\partial \Phi}{\partial y}\right|_{y=y_T} = -V_{0x} B_z(x) \tag{5}$$

where $V_{0x}$ is the plasma bulk velocity at the left hand side of the transition region. The boundary conditions (5) have been chosen such that the electric field at $y=y_B$ and $y=y_T$ sustains a quasi-uniform $\vec{E} \times \vec{B}$ drift in the $x$ direction: $E_y(x)/B_z(x) = V_{0x}$. The boundary conditions at $x=x_L$ and $x=x_R$ correspond to a vanishing $E_x$ component at the two sides. The electric field obtained from Eq. (4) and Eq. (5) is a two-dimensional generalization of the 1D electric field used in previous test-particle simulations [18]. Taking $x_R = -x_L$ and $y_T = -y_B$, the electric field intensity, $\vec{E} = -\nabla \Phi$, has the $E_x$ and $E_y$ components:

$$E_x(x,y) = -\frac{\pi}{2x_L} \sum_{m=1}^{\infty} \eta_m \cos\left(\frac{m\pi x}{2x_L}\right) \sinh\left(\frac{m\pi y}{2x_L}\right) \tag{6a}$$

$$E_y(x,y) = -\frac{\pi}{2x_L} \sum_{m=1}^{\infty} \eta_m \sin\left(\frac{m\pi x}{2x_L}\right) \cosh\left(\frac{m\pi y}{2x_L}\right) \tag{6b}$$

where $m$=1, 3, 5, etc. The $\eta_m$ coefficients are given by:

$$\eta_m = \frac{2x_L V_{0x} B_{1z}}{\pi^2 \cosh\left(\dfrac{m\pi y_B}{2x_L}\right)} \int_{-\pi}^{+\pi} \mathrm{erf}\left(\frac{2x_L}{\pi L}\theta\right) \sin(m\theta)\, d\theta \tag{7}$$

The solution of Laplace's equation (4) with boundary conditions (5) may be viewed as an electric field simulating the one sustained by space-charge layers forming at the boundaries of a moving non-diamagnetic plasma element in the presence of a magnetic field [19,20]. Our simulations have been performed for an electromagnetic field configuration that reproduces some typical parameters of the terrestrial magnetotail. The magnetic field profile would correspond to a tangential discontinuity. A possible origin of the electric field can be a localized perturbation of the dawn-dusk electric field. Another region of the magnetosphere where such electric and magnetic fields could be observed is the magnetopause. The relevance for this configuration of the electric and magnetic fields have been discussed in a previous publication [21].

The initial velocity distribution function specified for the source region is described by a displaced Maxwellian with the average velocity $\vec{V}_0$ parallel to the positive $x$-axis and perpendicular to the magnetic field:

$$f(v_x, v_y, v_z) = N_0 \left(\frac{m}{2\pi k_B T_0}\right)^{3/2} e^{-\frac{m\left[(v_x - V_0)^2 + v_y^2 + v_z^2\right]}{2k_B T_0}} \tag{8}$$

where $N_0$ and $T_0$ are the density and temperature of protons at the source region. In both forward and backward approaches the initial ($t$=0) source region, where the velocity distribution function is known, is localized in the left hand side of the transition region in the $xOy$ plane. It is defined in terms of the positions of the guiding centers.

In the forward approach a uniform grid of guiding centers having $N_x \times N_y$ nodes is placed inside the source region. For each of the $N_x \times N_y$ guiding centres, $n_p$ particles are "attached" with the initial velocities $(v_{x0}^i, v_{y0}^i, v_{z0}^i)$ distributed according to the displaced Maxwellian (8). Knowing the gyration velocity and the guiding center position of all test-particles, obtaining the particles' position is straightforward. In order to reconstruct the velocity distribution function at later times, $6 \times n_p \times N_x \times N_y$ equations of motion (1) are numerically integrated in the time range $t$>0, thus providing $3 \times n_p \times N_x \times N_y$ components of the test-particles velocities $(v_x^i, v_y^i, v_z^i)$ at time $t$. These final velocities define a scattered distribution of points in velocity space. Using the Liouville theorem (2) we assign to each point defined by the final velocities, $(v_x^i, v_y^i, v_z^i)$, the numerical value of the distribution function computed from (8) for the initial velocity components $(v_{x0}^i, v_{y0}^i, v_{z0}^i)$. In this way we obtain a map of $f$ in the three-dimensional velocity space. This procedure is applied at time $t$ in those spatial bins of the configuration space populated by a sufficiently large number of particles so as to have a good representation of $f$. A schematic diagram describing the forward approach is shown in Fig. 1.

With the backward approach, a three-dimensional velocity grid $(v_x^i, v_y^i, v_z^i)$ with $N_v$ vertices is constructed at a precise position in configuration space. Starting from each vertex of the velocity grid, the equation of motion of a test-particle is integrated backward in time back to $t$=0. To each node of the grid a single test-particle is assigned. In order to reconstruct the velocity distribution function at time $t$, $6 \times N_v$ equations of motion (1) are numerically integrated backward in time, thus providing $3 \times N_v$ components $(v_{x0}^i, v_{y0}^i, v_{z0}^i)$ of the test-particles velocities at time $t$=0. If the particle's $i$ guiding center is localized inside the source region, at time $t$=0, we assign to that particle the numerical value of the distribution function computed from (8) for the velocity components $(v_{x0}^i, v_{y0}^i, v_{z0}^i)$. Otherwise, the value of $f$ is set to zero. Using Liouville's theorem (2) we assign to each vertex of the grid, $(v_x^i, v_y^i, v_z^i)$, the numerical value of the distribution function assumed at time $t$=0. In this way $f$ is discretized in the three-dimensional velocity space. This procedure is applied at time $t$ for $N_r$ points of interest in configuration space. A schematic diagram describing the backward approach is shown in Fig. 2.

The numerical method used to solve the equation of motion of test-particles is based on the 4[th] order Runge-Kutta algorithm with fixed step-size. We checked the accuracy of the Runge-Kutta solver by integrating a number of 27 test-orbits with different initial conditions, forward and backward in time, over an interval of 225 seconds (~100 Larmor periods) using 1500 time steps. The results obtained show that the error in computing the particles positions is smaller than $0.13 R_L$, while the error in computing the particles velocities is smaller than $0.10 w_\perp$, where $R_L$ is the Larmor radius and $w_\perp$ is the gyration velocity of the particles. These results indicate that the accuracy of the numerical method used to integrate test-trajectories is satisfactory.

## 3. Numerical results and comparison between forward and backward test-kinetic approaches

The velocity distribution function is determined with the forward and backward test-kinetic approaches for different regions of a proton cloud moving in an anti-parallel magnetic field and a non-uniform electric field. The injection source is localized at the left hand side of the transition region and it is characterized by the displaced Maxwellian (8). The input parameters are given in Table I. The simulation domain is delimited by the interval [−40000, +40000] km along the $x$-axis and [−30000, +30000] km along the $y$-axis. The particles that reach regions outside these limits are removed from the simulation and no new particles are injected at the boundaries. In the forward approach, a uniform grid of 12×12 nodes considered as guiding centers is defined inside the source region; 20000 protons are "attached" to each guiding center position. The magnetic field, described by Eq. (3), is everywhere parallel to the $z$-axis and changes sign at $x$=0 (see Fig. 3 – left panel). Figure 3 (right panel) shows the electric field profile obtained from the Laplace equation (4) subject to boundary conditions (5) discussed in the previous section. This profile may be viewed as describing a neutral sheet and a superimposed electric field with $E_y$ changing sign whenever $B_z$ reverses sign.

Figure 4 (left panel) shows the initial positions of protons ($t$=0) and the local number density in the $xOy$ plane, perpendicular to the magnetic field. A two-dimensional cross-section (for $v_z$=0) of the velocity distribution function corresponding to the central region (the blue rectangle in the $xOy$ plane) is shown in the right panel of Fig. 4. It is shown that the initial distribution function is a displaced Maxwellian with an average velocity $V_0$=200 km/s in the $x$ direction.

The positions and the local number density of the test-protons in the $xOy$ plane, perpendicular to the magnetic field, at $t$=225 s (~100·$T_L$) are shown in Fig. 5. The overall shape of the proton cloud is deformed and shows significant asymmetries, the particles being scattered in the positive direction of the $y$-axis. This asymmetric expansion of the cloud is related to the gradient-B drift that is oriented in the +$Oy$ direction. Thus, an energy-dispersed structure is formed due to the energy-dependent displacement of protons towards the edges of the cloud by the gradient-B drift. Another effect of the gradient-B drift is the formation of ring-shaped velocity distribution functions within the energy-dispersed structure, as can be seen further. Higher energy particles populate the edges of the proton beam while smaller energies are located inside the core. Also, non-gyrotropic velocity distribution functions form in the front-side and trailing edge of the cloud due to remote sensing of energetic particles with guiding centers localized inside the beam. An explanation of the physical mechanism responsible for the formation of such an energy-dispersed structure and also a detailed analysis of the kinetic effects contributing to the formation of ring-shaped and non-gyrotropic velocity distribution functions is published elsewhere [21]. Here we limit our attention to the differences between the results obtained using both forward and backward test-kinetic approaches.

The velocity distribution function of protons obtained at $t$=225 s using the forward approach is shown in Fig. 6. The velocity distribution function inside the cloud is computed for each bin defined by the blue rectangles in the $xOy$ plane and identified by the combination of letters (columns) and numbers (rows) in Fig. 5. The size of a spatial bin is defined such that it contains enough particles (at least $10^4$ particles per bin) for a good sampling of velocity space. The bins of the mesh shown in Fig. 5 have a spatial resolution of 280 km in $x$-direction and 2500 km in $y$-direction, adapted to the geometry of the cloud and the total number of simulated particles. The corresponding velocity distribution functions obtained using the backward approach are shown in Fig. 7. $f$ is computed for the central point of each spatial bin defined by the blue rectangles in the $xOy$ plane illustrated in Fig. 5.

There are significant differences between the forward and backward approaches as illustrated by Fig. 6 and Fig. 7. Nevertheless, the velocity distribution function, $f$, has the same variation tendency in both cases, i.e. (i) it is ring-shaped close to the upper boundary (i.e. for larger $y$-values) while in the center is approximately Maxwellian (comparing, for instance, $f$ corresponding to C1 and C3 in Fig. 6 and Fig. 7) and (ii) the anisotropy of the velocity distribution function is more pronounced close to the trailing edge of the cloud (i.e. for smaller $x$-values) than in the center (for example, comparing $f$ for A2 and B2 in Fig. 6 and Fig. 7). The explanation for the differences observed is related to the different methods used to compute the distribution function with forward and backward approaches. In the forward approach $f$ is sampled over a spatial bin whose size is defined such that it contains a large enough number of particles and the statistical error resulting from sampling is minimized. On the other hand, in the backward approach the computation of $f$ for a precise point in configuration space is free from statistical errors; in our case, $f$ is computed for the central point of each spatial bin defined for the forward method. The strength of the backward approach is related to its ability to produce detailed velocity distribution functions at precise locations without statistical sampling errors. The essential difference between the forward and backward approaches is that the former necessarily relies on spatial binning and sampling, while the latter can be calculated at precise locations in space. In many cases, the backward approach can lead to filamentary structures in velocity (or momentum) space, while such structures are always attenuated in the forward approach, owing to the spatial averages involved. In contrary to the backward approach, the forward Liouville approach enables the computation of both the velocity distribution function and general dynamics of the particle cloud while advancing an initial distribution of particles into a non-uniform configuration of the magnetic and electric fields. Thus, the strength of the forward approach is related to its ability to investigate the evolution of a specific plasma source.

A solution to eliminate these differences and to obtain comparable distribution functions would involve spatial averages of $f$ by a proper quadrature scheme applied in the backward approach. For that purpose, the velocity distribution function obtained using the backward approach is numerically integrated over a rectangular domain in the $xOy$ plane corresponding to the spatial bin used to compute the distribution function using the forward approach. The resulting averages are presented in Fig. 8 for each bin defined by the blue rectangles in the $xOy$ plane (see Fig. 5). The averages are computed by the trapezoidal integration rule with 10×10 points applied in each spatial bin. The resulting averaged distribution functions are closer to those given by the forward approach, as expected. Nevertheless, there are still some notable differences particularly for bins B2 and C2 (see Fig. 8). These two bins are localized in a region characterized by a pronounced spatial variation of the velocity distribution function, as can be seen from Fig. 7 by comparing $f$ for C1* and C2 (C1* is the middle point between C1's and C2's centres). On the other hand, the results obtained for bin C1 using both forward and backward approaches are very similar since the spatial variation of the distribution function for that region is smooth (see $f$ for C1 and C1* in Fig. 7). The differences observed for the bins localized in regions with sharp spatial variation of the velocity distribution function can be explained by analyzing in more detail the sampling method of the forward approach and the averaging method of the backward approach. In order to better understand the differences between the two, a schematic representation is shown in Figures 9 and 10.

Let us consider the problem of calculating the velocity distribution function for a spatial bin which is localized in a region from the configuration space characterized by a steep spatial variation of $f$ along $Oy$ direction. The velocity distribution function is computed using both forward and backward approaches; for the latter approach, the averaging method is used. We divide the spatial bin in two areas, A and B, characterized by two distinct velocity distribution functions, as shown in Fig. 9 and 10. In area A the velocity distribution function is the inner core of a Maxwellian distribution function, $f_A$, while in area B we retrieve the outer shell of the same Maxwellian distribution, $f_B$.

For simplicity, let us assume that $f$ does not vary significantly in either A or B. By using the forward approach, the particles localized in the both areas A and B will have the velocities distributed according to their respective velocity distribution functions. Thus, the less energetic particles will be found in area A, while the most energetic ones will be found in area B (see Fig. 9). This simplified model corresponds roughly to our simulation results. All particles localized inside the entire spatial bin will be distributed in velocity space as follows. Particles originating from area A, i.e. the less energetic ones, will be found in the central regions of velocity space, while particles originating from area B, i.e. the most energetic ones, will be found in the outer regions of velocity space.

In order to reconstruct the velocity distribution function by using the forward approach, a uniform grid in velocity space is defined. For each velocity bin $j$ centred in $\vec{v}_j$, the corresponding distribution function $f_{\mathrm{FWD}}(\vec{v}_j)$ is computed by averaging over all numerical values $f_j^i$ "attached" to each particle $i$ localized inside the considered velocity bin:

$$f_{\mathrm{FWD}}(\vec{v}_j) = \frac{\sum_{i=1}^{n_j} f_j^i(\vec{v}_j^i)}{n_j} \tag{9}$$

where $\vec{v}_j^i$ is the velocity of particle $i$ localized inside the velocity bin $j$ and $n_j$ is the total number of particles inside bin $j$. Among these $n_j$ particles, let $n_j^A$ be the ones from area A and $n_j^B$ those from area B such that $n_j = n_j^A + n_j^B$. Thus, Eq. (9) becomes:

$$f_{\mathrm{FWD}}(\vec{v}_j) = \frac{\sum_{i_A=1}^{n_j^A} f_j^{i_A}(\vec{v}_j^{i_A}) + \sum_{i_B=1}^{n_j^B} f_j^{i_B}(\vec{v}_j^{i_B})}{n_j} \tag{10}$$

where $f_j^{i_A}(\vec{v}_j^{i_A}) = f_A(\vec{v}_j^{i_A})$, since all $i_A$ particles are localized inside area A and likewise $f_j^{i_B}(\vec{v}_j^{i_B}) = f_B(\vec{v}_j^{i_B})$, as all $i_B$ particles belongs to area B. Furthermore, we consider that all velocity bins are small enough such that $f_A(\vec{v}_j^{i_A}) = f_A(\vec{v}_j)$ and $f_B(\vec{v}_j^{i_B}) = f_B(\vec{v}_j)$. In this way, $f$ computed using the forward approach for the velocity bin centred on $\vec{v}_j$ is:

$$f_{\text{FWD}}(\vec{v}_j) = \frac{n_j^A}{n_j} f_A(\vec{v}_j) + \left(1 - \frac{n_j^A}{n_j}\right) f_B(\vec{v}_j) \tag{11}$$

In order to compute the velocity distribution function using the backward approach, we define a uniform grid in configuration space having $n \times n$ points that cover the entire area of the spatial bin to be sampled (see Fig. 10). For each velocity vertex $j$ centred in $\vec{v}_j$, the corresponding distribution function $f_{\text{BWD}}(\vec{v}_j)$ is computed by averaging over all numerical values $f_j^i$ "attached" to each point $i$ of the spatial grid:

$$f_{\text{BWD}}(\vec{v}_j) = \frac{\sum_{i=1}^{n^2} f_j^i(\vec{v}_j)}{n^2} \tag{12}$$

Considering that $m$ grid points are localized inside area A, while the other $n^2-m$ grid points are localized inside area B, Eq. (12) becomes:

$$f_{\text{BWD}}(\vec{v}_j) = \frac{\sum_{i_A=1}^{m} f_j^{i_A}(\vec{v}_j) + \sum_{i_B=m+1}^{n^2} f_j^{i_B}(\vec{v}_j)}{n^2} \tag{13}$$

where $f_j^{i_A}(\vec{v}_j) = f_A(\vec{v}_j)$, since all $i_A$ grid points are localized inside area A, and similarly $f_j^{i_B}(\vec{v}_j) = f_B(\vec{v}_j)$. Therefore, $f$ computed using the backward approach for the velocity bin centred on $\vec{v}_j$ is given by:

$$f_{\text{BWD}}(\vec{v}_j) = \frac{m}{n^2} f_A(\vec{v}_j) + \left(1 - \frac{m}{n^2}\right) f_B(\vec{v}_j) \tag{14}$$

We should mention that the average value (14) has been obtained simply by computing the arithmetic mean of all $n^2$ function's values instead of integrating the velocity distribution function over the entire spatial bin using a 2D trapezoidal integration rule, as it is done in our simulations. Also, we considered a uniform grid in velocity space for the backward approach, while in our simulations an unstructured grid has been used to compute the velocity distribution function. These simplifications should not have major consequences on the final results.

In order to compare the velocity distribution functions obtained from both forward and backward approaches we considered three representative velocity bins, designated $a$, $b$ and $c$ and centred at $\vec{v}_a$, $\vec{v}_b$ and $\vec{v}_c$ (see Fig. 9 and Fig. 10), to compute the numerical values of $f$ given by Eq. (11) and Eq. (14). These three velocity bins have been chosen such that:

$f_A(\vec{v}_a) \neq 0$ and $f_B(\vec{v}_a) = 0$, $f_A(\vec{v}_b) = f_B(\vec{v}_b) \neq 0$, while $f_A(\vec{v}_c) = 0$ and $f_B(\vec{v}_c) \neq 0$. Therefore, from Eq. (11) and Eq. (14), we obtain the values of $f$ computed with both forward and backward approaches for velocity bins $a$, $b$ and $c$. The results are given in Table II and show that the velocity distribution function given by the backward approach is smaller than the one obtained from the forward approach for velocity bins $a$ and $c$, while for bin $b$ both values are equal.

By applying this algorithm to all velocity space bins, a Maxwellian distribution function is obtained with the forward approach, as can be seen in Fig. 9. However, $f$ obtained with the backward approach presents a cavity in the central region of velocity space, as can be seen in Fig. 10. Similar results are obtained, for instance, for bin C2 of our simulations depicted in Fig. 6 and Fig. 8, which is localized in a region characterized by a steep spatial variation of the velocity distribution function. Indeed, with the forward approach a Maxwellian distribution is obtained for bin C2, while with the backward approach the distribution function is characterized by a central cavity in velocity space. Thus, the simplified model described in Fig. 9 and Fig. 10 explains the differences obtained between forward and backward approaches in spatial regions characterized by sharp gradients of $f$.

The velocity distribution functions given by Eq. (11), for the forward approach, and Eq. (14), for the backward approach, have similar mathematical expressions except for the weight coefficients of $f_A$ and $f_B$. In the forward approach the weight coefficients are expressed in terms of $n_j^A / n_j$, i.e. the ratio of the number of particles localized inside velocity bin $j$ and pertaining to spatial area A to the total number of particles localized inside velocity bin $j$. In the backward approach the weight coefficients are expressed in terms of $m / n^2$, i.e. the ratio of grid points number localized inside area A to total number of grid points localized inside the entire spatial bin. By analyzing the $n_j^A / n_j$ ratio we can conclude that this quantity depends on the position of bin $j$ in velocity space. On the other hand, $m / n^2$ is equal to the ratio of region A area to entire spatial bin area, which is independent on the position of bin $j$ in velocity space:

$$\frac{m}{n^2} = \frac{Area(A)}{Area(bin)} = \frac{L_y^A}{L_y} \tag{15}$$

where $L_y^A$ indicates the width of area A along the $y$-axis, while $L_y$ represent the width of the entire spatial bin. Thus, the weight coefficients corresponding to forward and backward distribution functions are not equal in general and the results provided by the two approaches may also be different, independently of the number of particles injected in the forward simulations or the number of grid points used in the averaging scheme for the backward simulations.

The main point which distinguishes the averaging method (14) from the sampling method (11) is related to the fact that, in the backward approach, to a given point in velocity space correspond $n^2$ points in the configuration space which cover the entire area of the spatial bin. In the forward approach however, a given bin in velocity space may originate from only a subset of points in configuration space localized in a certain area of the spatial bin. Therefore, in order to calculate the numerical value of the distribution function at a certain bin in velocity space, the backward averaging method (14) will take into account the contribution from the entire spatial bin, while the forward sampling method (11) will take into account the contribution of only a part of the considered spatial bin, thus possibly leading to different results. Nevertheless, $f_{\text{FWD}}$ given by Eq. (11) would be equal to $f_{\text{BWD}}$ given by Eq. (14) if $m = n^2$. This condition is satisfied if we increase the size of region A such that it will cover the entire area of the spatial bin. Only in this case $n_j^A$ will also be equal to $n_j$ for all velocity space bins and the weight coefficients corresponding to forward and backward distribution functions will be equal. Therefore, by increasing the size of area A it is possible to obtain converging results with both approaches as long as the initial assumption is satisfied, i.e. there are no significant spatial variations of $f$ along area A. We should note that this assumption will always be satisfied for region B since the size of this area continually decreases, as the size of A increases. This result can be generalized for three-dimensional bins with spatial variations of the velocity distribution function along all three coordinate axes. In this case the forward and backward approaches will return similar results only for those spatial bins which are small enough such that the following inequality to be satisfied simultaneously along all three coordinates axes:

$$L_i \cdot \left| \frac{\partial f}{\partial x_i} \right| << f \tag{16}$$

where $i = 1, 2, 3$ for the $x$, $y$, $z$ axes respectively. On the other hand, with bins covering regions of configuration space characterized by sharp spatial gradients of the velocity distribution function, the forward and backward approaches will generally provide different results.

## 4. Conclusions

In this paper we performed a comparative study of the forward and backward Liouville approaches corresponding to the test-kinetic simulation method that integrates numerically test-particle orbits in given electric and magnetic fields. The test-kinetic method has been applied to study various problems of space plasma physics. In this paper we discuss an example relevant for magnetospheric physics that is analyzed in detail in a previous publication [21]. The test-kinetic method is an important simulation tool, especially in complex situations where the use of fully self-consistent kinetic methods is not possible. We applied the forward and backward approaches to compute the velocity distribution function in different areas of a proton cloud moving in the vicinity of a region with a sharp transition of the magnetic field and a non-uniform electric field. The source region is localized in the left hand side of the transition region and it is characterized by a displaced Maxwellian distribution function.

We compare the velocity distribution functions obtained for different regions of the proton cloud with the forward and backward approaches. In the forward approach $f$ is sampled over a spatial bin which needs to be populated by a sufficiently large number of particles so as to reduce statistical errors. On the other hand, in the backward approach $f$ is computed without statistical errors, at precise positions in configuration space. In order to compare the distribution functions obtained with both approaches, a spatial averaging of $f$ is needed. The velocity distribution function given by the backward approach is numerically integrated over a rectangular domain corresponding to the spatial bin used to compute the distribution function with the forward approach.

Our simulation results show that there are significant differences between the distribution functions given by forward and backward approaches. The differences are observed especially for spatial bins from regions with a steep spatial variation of the velocity distribution function, while in regions with smooth variations of $f$ the two approaches provide similar results. The differences and similarities can be explained by a careful examination of the sampling method used in the forward approach and the averaging method used in the backward approach. The main difference between the two computational methods is due to the approach used to estimate the velocity distribution function in a spatial bin: the backward method uses an averaging method that takes into account the contribution of the entire spatial bin to calculate the distribution function for a certain bin in velocity space, while, in certain cases, the forward sampling method effectively only takes into account the contribution from a part of the bin considered. The two approaches lead to similar results when averages are calculated over bins in which the distribution function varies smoothly in configuration space.

**Acknowledgements**

This paper is supported by the Sectoral Operational Programme Human Resources Development (SOP HRD) financed from the European Social Fund and by the Romanian Government under Contract No. SOP HRD/107/1.5/S/82514. Marius Echim and Gabriel Voitcu acknowledge support from the European Space Agency (ESA) through PECS Project no. 98049/2007 − KEEV and thanks International Space Science Institute in Bern, Switzerland for supporting the team "Plasma entry and transport in the plasma sheet" led by Simon Wing. Richard Marchand was supported by the Natural Sciences and Engineering Research Council of Canada.

Table I. Input parameters of the test-kinetic simulations: $N_0$, $k_B T_0$, $V_0$ are the density, thermal energy and average velocity of the displaced Maxwellian given in Eq. (8). Here $B_{1z}$ is the asymptotic value of the magnetic field at the left hand side of the transition region, $L$ is the length scale of the transition region, $R_L$ is the Larmor radius of thermal protons and $T_L$ is the Larmor period of protons at the left hand side of the transition region. The boundaries of the source region in the $xOy$ plane are defined by $x_{min}$, $x_{max}$, $y_{min}$, $y_{max}$.

| $N_0$ | $k_B T_0$ | $V_0$ | $B_{1z}$ | $L$ | $R_L$ | $T_L$ | $[x_{min}\ x_{max}] \times [\ y_{min}\ y_{max}]$ |
|---|---|---|---|---|---|---|---|
| $[m^{-3}]$ | $[eV]$ | $[km/s]$ | $[nT]$ | $[km]$ | $[km]$ | $[s]$ | $[km]$ |
| $10^4$ | 3000 | 200 | $-30$ | 6000 | 260 | 2.2 | $[-20000, -17800] \times [-550, +550]$ |

Table II. Values of $f$ obtained with both forward and backward approaches for three selected velocity bins centered at $a$, $b$ and $c$.

| $\vec{v}_j$ | $f_{\text{FWD}}(\vec{v}_j)$ | $f_{\text{BWD}}(\vec{v}_j)$ |
|---|---|---|
| $\vec{v}_a$ | $f_A(\vec{v}_a)$ | $\dfrac{m}{n^2} f_A(\vec{v}_a) < f_A(\vec{v}_a)$ |
| $\vec{v}_b$ | $f_A(\vec{v}_b)$ $f_B(\vec{v}_b)$ | $f_A(\vec{v}_b)$ $f_B(\vec{v}_b)$ |
| $\vec{v}_c$ | $f_B(\vec{v}_c)$ | $\left(1 - \dfrac{m}{n^2}\right) f_B(\vec{v}_c) < f_B(\vec{v}_c)$ |

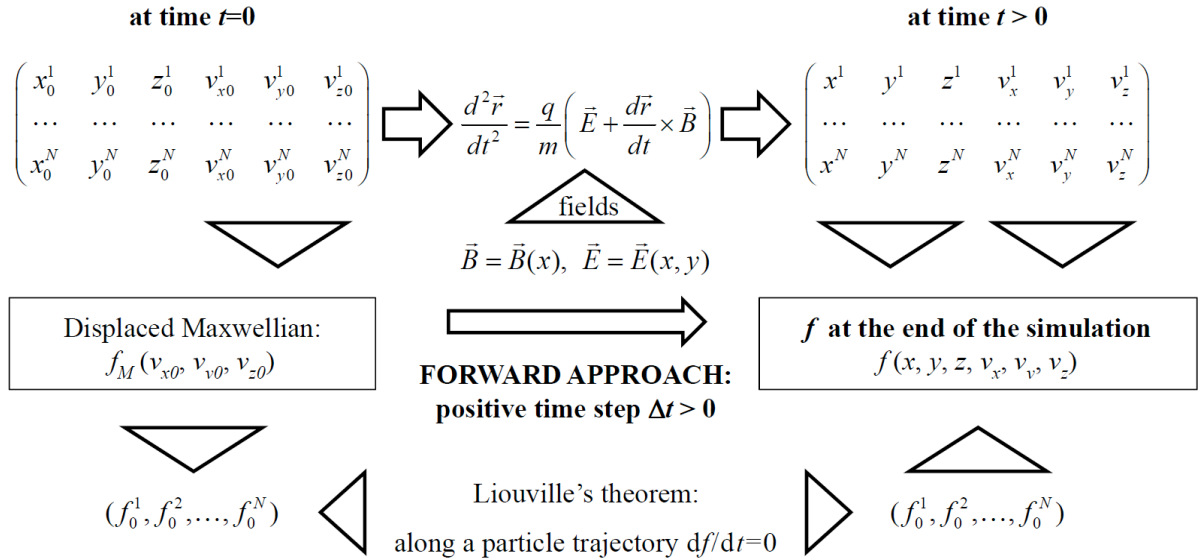

Figure 1: Schematic diagram of the forward Liouville approach; a positive time step is used to integrate test-particle orbits in given magnetic and electric fields.

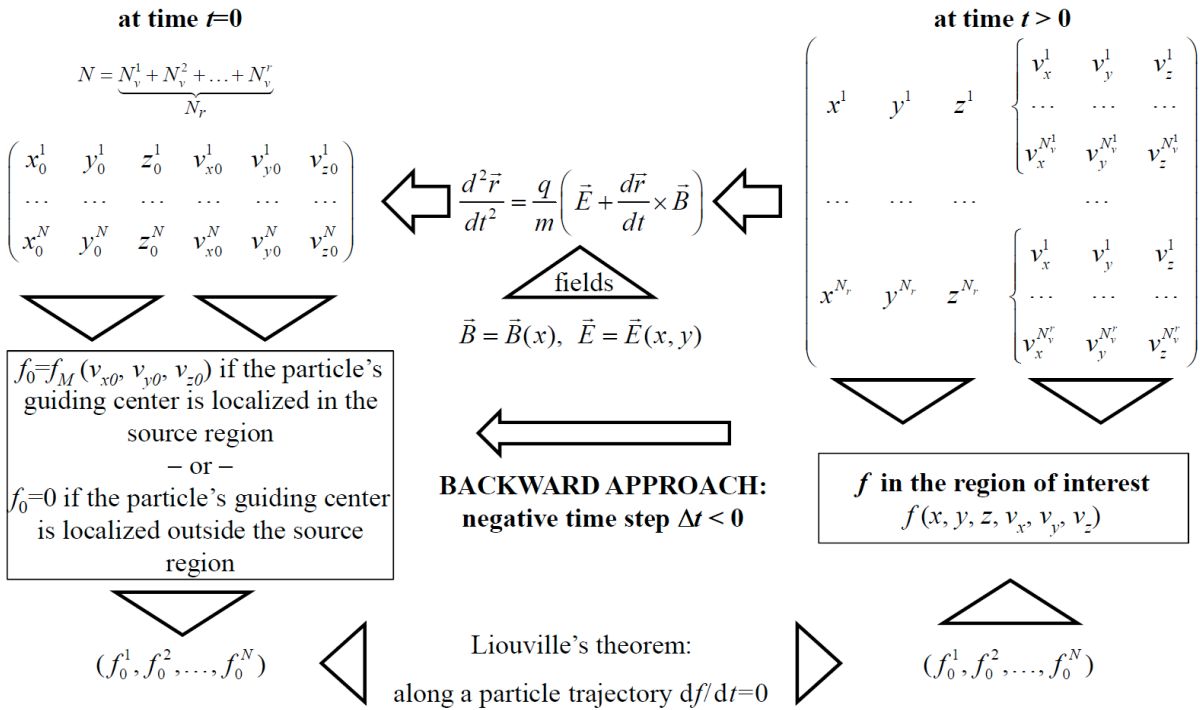

Figure 2: Schematic diagram of the backward Liouville approach; a negative time step is used to integrate test-particle orbits in given magnetic and electric fields.

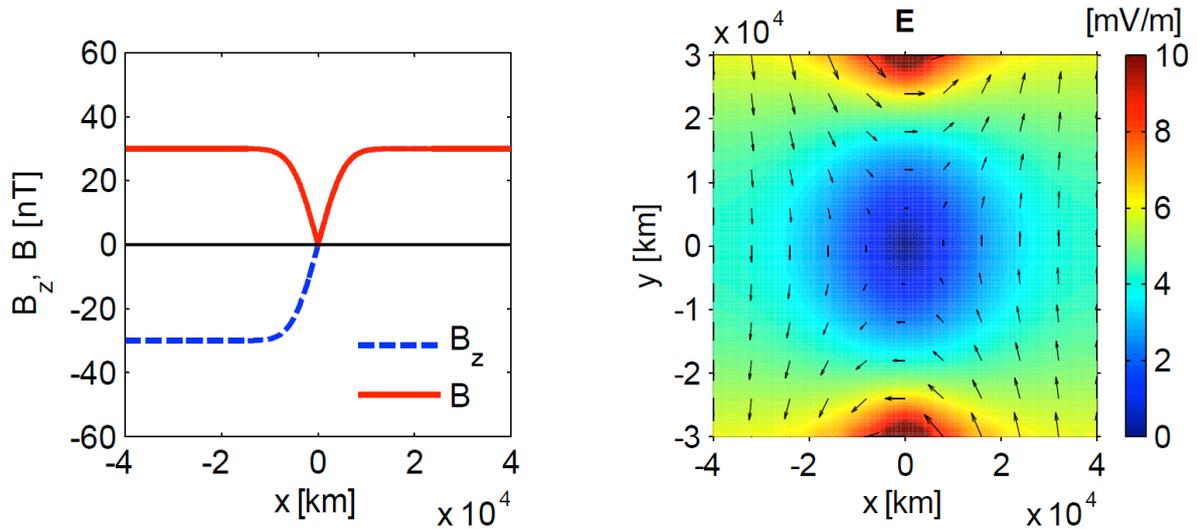

Figure 3: Left panel: magnetic field profile in the simulation domain; the B-field is unidirectional and changes orientation at $x=0$. Right panel: electric field profile in the simulation domain; $E_y$ changes sign whenever $B_z$ reverses sign. The simulation domain is limited by: $-40000$ km $\leq x \leq +40000$ km, $-30000$ km $\leq y \leq +30000$ km.

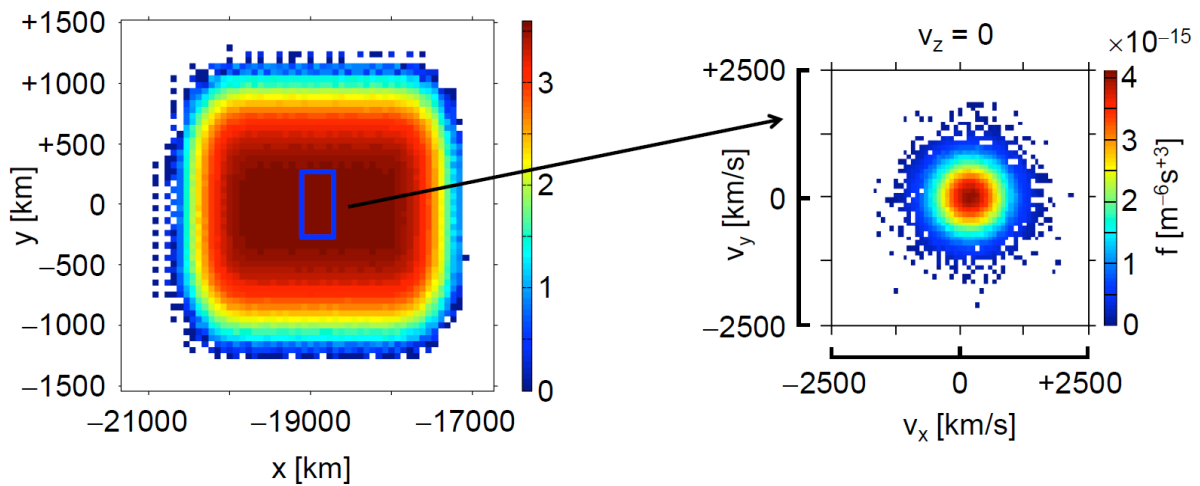

Figure 4: The left panel shows proton density profile in the $xOy$ plane, perpendicular to the magnetic field, at $t=0$. The local number density is color coded using a 2D mesh of 60x60 spatial cells. The right panel shows the velocity distribution function, in the $v_z=0$ cross section, sampled in the blue rectangle in the left panel. Note that the initial distribution function is given by a displaced Maxwellian with an average velocity $V_0=200$ km/s along the $x$ direction.

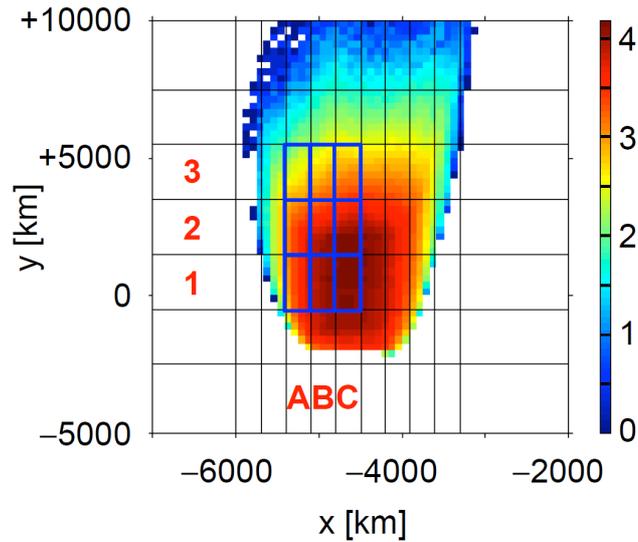

Figure 5: Proton density distribution in the *xOy* plane, perpendicular to the magnetic field, at *t*=225s (~100·*T<sub>L</sub>*) obtained with the forward approach. The local number density is color coded using a 2D mesh of 60x60 spatial cells. The density distribution is elongated in +*Oy* direction due to the gradient-B drift acting in the region of non-uniform fields. The blue rectangles indicate the spatial bins used to sample the velocity distribution function shown in Fig. 6.

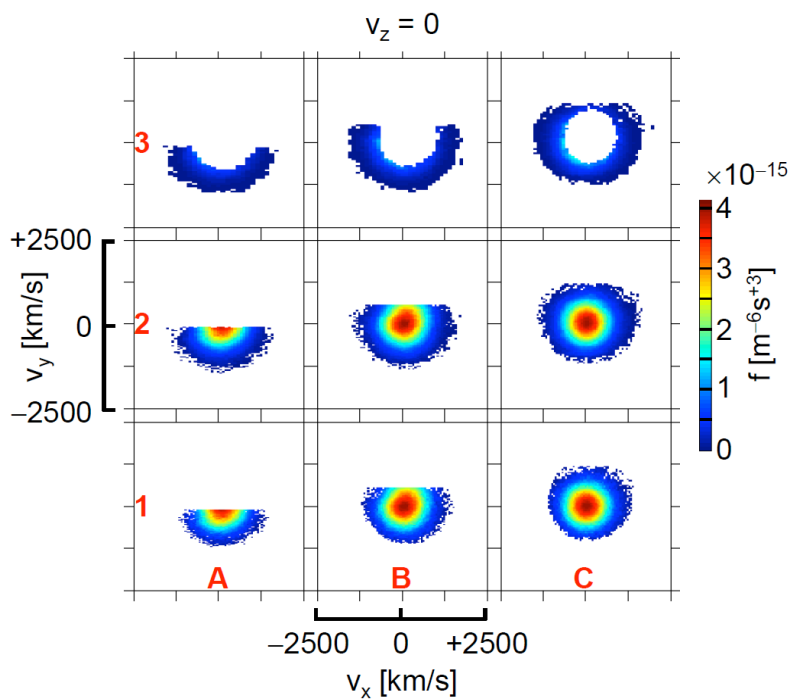

Figure 6: Velocity distribution functions obtained at *t*=225 s (~100·*T<sub>L</sub>*) using the forward approach in the spatial bins indicated by blue rectangles in Fig. 5. The plots correspond to *v<sub>z</sub>*=0 cross-sections in velocity space.

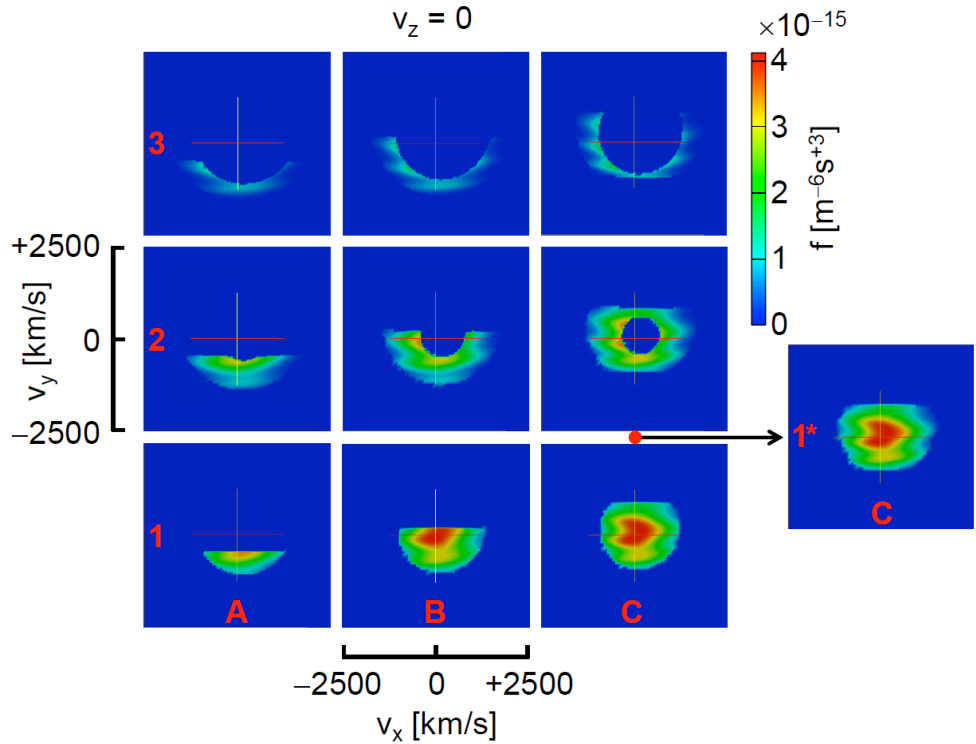

Figure 7: Velocity distribution functions obtained at $t$=225 s (~100·$T_L$) using the backward approach for the central points of the bins indicated by blue rectangles in Fig. 5. C1* is the middle point between C1 and C2. The plots show $v_z$=0 cross-sections in velocity space.

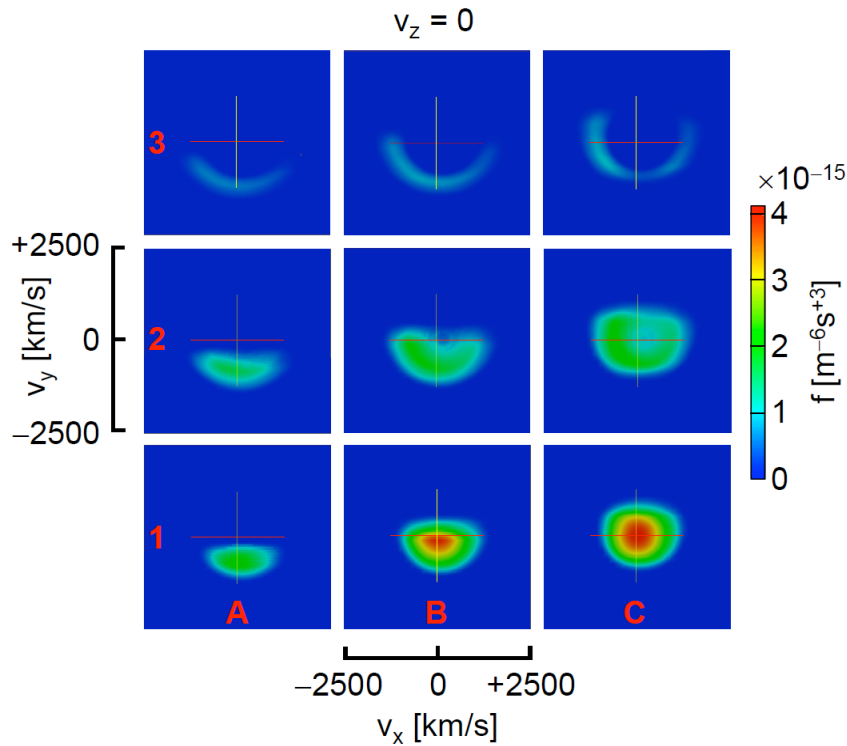

Figure 8: Velocity distribution functions obtained at $t$=225 s (~100·$T_L$) with the backward approach by averaging over the spatial bins indicated with blue rectangles in Fig. 5. The plots show $v_z$=0 cross-sections in velocity space.

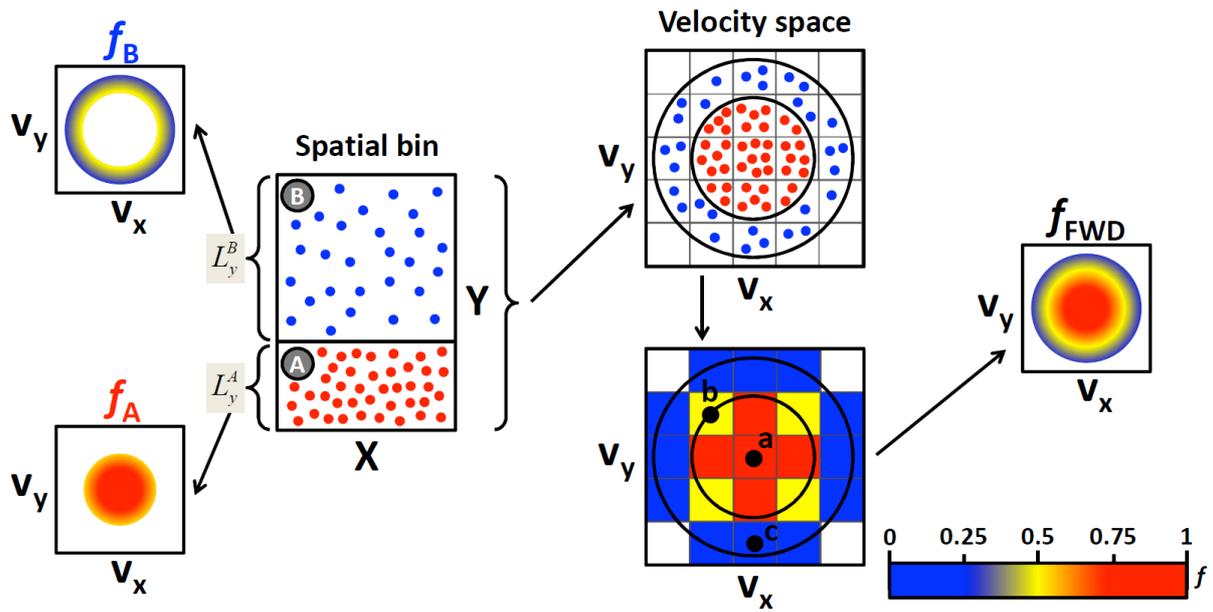

Figure 9: Schematic diagram illustrating the sampling method used to compute the velocity distribution function with the forward Liouville approach.

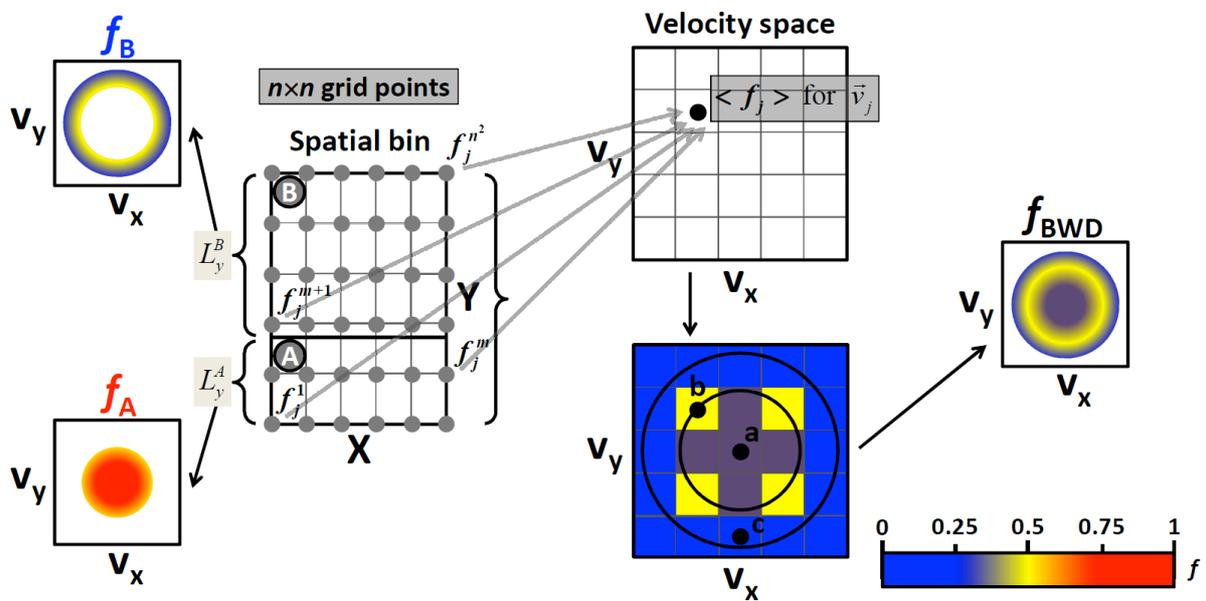

Figure 10: Schematic diagram illustrating the averaging method used to compute the velocity distribution function with the backward Liouville approach.